\documentclass{aa}
\usepackage{natbib,array,multirow,epsfig,graphicx,amssymb,amsmath,lscape}
\usepackage{txfonts,hyperref,rotating,psfrag}
\usepackage{aas_macros}

\begin{document}

\title{New electron-proton Bremsstrahlung rates for a hot plasma where the electron
  temperature is much smaller than the proton temperature}
\titlerunning{Bremsstrahlung in a two-temperature plasma}
\author{M. Mayer\inst{1}}
\institute{Institute of Astronomy, Madingley Road,
Cambridge CB3 0HA, UK\\
\email{mm@ast.cam.ac.uk}}

\date{Received \today; accepted sometime}
\newcommand{\cb}{ }
\newcommand{\cm}{\textrm{ cm}}
\newcommand{\g}{\textrm{ g}}
\newcommand{\K}{\textrm{ K}}
\newcommand{\jw}{\textrm{j/w}}
\date{\today }

\bibliographystyle{aa}

\abstract
{Observations of X-Ray sources harbouring a black hole and an
  accretion disc show the presence of at least two spectral components. One
  component is black-body radiation from an optically thick
  standard accretion disc. The other is produced in a optically thin
  corona and usually shows a powerlaw behaviour. Electron-proton (ep) bremsstrahlung
  is one of the contributing radiation mechanisms in the corona. Soft
  photons from the optically thick disc can Compton cool the electrons
in the corona and therefore lead to a two-temperature plasma, where
electrons and ions have different temperatures.}
{We qualitatively discuss effects on ep-bremsstrahlung in the presence
  of such a two-temperature plasma.}
{ We use the classical dipole
approximation allowing for non-relativistic electrons and protons and apply quantum corrections through high-precision
Gaunt factors. }
{In the two-temperature case ($T_\textrm{e}< T_\textrm{p}$) the protons
 cause
a significant fraction of the ep-bremsstrahlung if their speed is high
compared to the electrons. We give accurate values for
ep-bremsstrahlung including quantum-mechanical corrections in the non-relativistic limit and give some
approximations in the relativistic limit.}
{The formulae presented in this paper can be used in models of
  black hole accretion discs where an optically thin corona can
  comprise a two-temperature plasma. This work could be extended to
  include the fully relativistic case if required.}


\maketitle

\section{Introduction}

Models of black hole accretion discs are widely used to explain
spectral characteristics of X-Ray observations of such objects. 
These models usually consist of an optically thin and hot corona and
a cool, optically thick standard accretion disc. The
optically thin corona consists of a two-temperature plasma. Electrons
cool by bremsstrahlung, mainly by electron-proton bremsstrahlung. 
The electron-proton system has got a dipole moment while the
electron-electron and proton-proton system can only emit
radiation through the (much weaker) quadrupole moment.

\citet{1975ApJ...199L.153E} presented a model for Cyg
X-1 incorporating a two-temperature plasma in the inner parts of the
accretion flow. This kind of model has been subsequently developed
\citep[e.g.][]{1976ApJ...204..187S,1995MNRAS.277...70Z} and applied to
different objects \citep{1991ApJ...380L..51H,1993ApJ...413..507H} and
different geometries \citep{1991ApJ...380...84W}.

The calculation of the rate of bremsstrahlung coming from a hot gas 
has received much attention over a long period
of time
\citep[e.g.][]{1962RvMP...34..507B,1970RvMP...42..237B,1980ApJ...238.1026G,1981ApJ...243..677G}.
Today the bremsstrahlung rate is known with a very high accuracy. 
All these calculations however assume electron and proton temperature to be
equal. 

In the following we limit ourselves to a purely classical and
non-relativistic treatment of electron-proton bremsstrahlung. We do
this in order to highlight the elementary physical processes
involved. Since we apply Gaunt factors to account for quantum
mechanical effects, our results are strictly accurate only in the
non-relativistic case. An expansion to the fully relativistic case
is feasible but beyond the scope of this paper.

In this contribution we first highlight the physical mechanism in
Sect.~\ref{sect:mechanism} and then present a recalculation of the 
electron-proton Bremsstrahlung rate for a
two-temperature plasma in the non-relativistic limit in
Sect.~\ref{sect:calc}. While we limit ourselves to a electron-proton
plasma, the formulae can easily be modified to consider mixtures of
ions with different mass and charge. We give a discussion of the results and our conclusions in Sect.~\ref{sect:conclusions}.

\section{The mechanism} \label{sect:mechanism}

\begin{figure}
  \centering
  \includegraphics[width=0.45\textwidth]{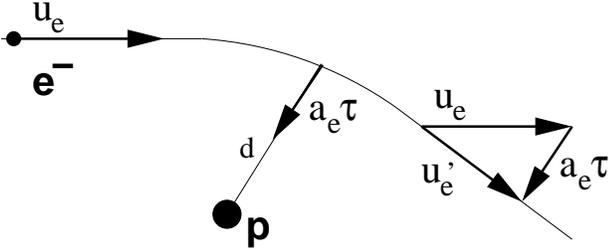}
  \caption{Trajectory of an electron ($\cb e^-$) at speed $\vec u_\textrm{e}$. For
    simplicity the proton ($\cb p$) is assumed to be at rest. At the closest
    approach, the acceleration $\vec a_\textrm{e}\tau$ moves the electron
    closer to the proton and the electron loses kinetic energy.}
  \label{fig:pic}
\end{figure}

Electron-proton Bremsstrahlung in the standard picture is created through the acceleration of electrons in the
field of a proton. This acceleration leads to the emission of
electromagnetic radiation resulting in the loss of kinetic electron energy. We show this situation in Fig.~\ref{fig:pic}.

Consider a electron and a proton travelling at a speed $\vec u_\textrm{e}$ and
$\vec u_\textrm{p}$, respectively, have an encounter at a minimal distance
$d$ and at a relative speed of $\vec u_{rel}=\vec u_\textrm{e}-\vec u_\textrm{p}$. 
The Coulomb field leads to a mutual attraction with the
corresponding Coulomb force $\vec F_\textrm{p}=-\vec F_\textrm{e}$ with $\vec
F_\textrm{e}\propto d^{-2}$ and subsequently to
an acceleration $\vec a_\textrm{p}=\vec F_\textrm{p}/m_\textrm{p}$ and $\vec a_\textrm{e}=\vec
F_\textrm{p}/m_\textrm{e}$. This interaction happens on a characteristic timescale
$\tau\approx d/u_{rel}$ and the change in the speed of the particles
is small compared to their velocity (small-angle scattering). 

After the interaction, the speed of the electron is $\vec u_\textrm{e}'=\vec
u_\textrm{e}+\vec a_\textrm{e}\tau $ and that of the proton is $\vec u_\textrm{p}'=\vec u_\textrm{p}+\vec
a_\textrm{p}\tau$. The change in kinetic energy for the electron and proton can
be written as 
 \begin{eqnarray}
  \Delta E_{kin,e}&=& m_\textrm{e} \tau \vec a_\textrm{e} \cdot \vec u_{e} \label{eg:kinenergye}\;,\\
  \Delta E_{kin,p}&=&-m_\textrm{e} \tau \vec a_\textrm{e} \cdot \vec u_{p} \label{eg:kinenergyp}\;,
\end{eqnarray}
where we have made use of the fact that $\vec a_\textrm{p}=-m_\textrm{e}/m_\textrm{p}\vec
a_\textrm{e}$ and the change in speed $|\Delta \vec u|<<|a\tau|$. Thus we can
neglect terms of the order $\tau^2$. Both the
electrons and the proton lose kinetic energy ($\vec a_\textrm{e}$ and $\vec
u_\textrm{e}$ form an angle larger than $\pi/2$ and thus their scalar product
is negative). 

It is evident from eqns.~(\ref{eg:kinenergye}) and (\ref{eg:kinenergyp})
that for a one-temperature plasma, where the mean speed of the
electron is faster by a factor of
$\left(m_\textrm{p}/m_\textrm{e}\right)^{1/2}\approx 43$ than the speed of the
protons, most of the energy of the electron-proton
bremsstrahlung arises from the kinetic energy change of the electrons.

However for a two-temperature plasma the ratio for the mean speed of
protons and electrons is given by
\begin{equation}
  \label{eq:speedratio}
  \frac{u_\textrm{e}}{u_\textrm{p}}=\sqrt{\frac{T_\textrm{e}}{T_\textrm{p}}\frac{m_\textrm{p}}{m_\textrm{e}}}\;.
\end{equation}
For $T_\textrm{p}/T_\textrm{e}>m_\textrm{p}/m_\textrm{e}\approx 1836$ the mean speed of the protons
exceeds the mean speed of the electrons and then the change in kinetic
energy is greater for the protons. In this case it is the protons
which are responsible for providing the energy to create a photon leading to
electron-proton bremsstrahlung.

We now explore this mechanism in a classical and non-relativistic
treatment of electron-proton bremsstrahlung in the dipole-approximation.

\section{Non-relativistic electron-proton Bremsstrahlung} \label{sect:calc}

The emission per unit time $t$, volume $V$ and photon energy  $E_\nu=h \nu$ for
a single proton can be expressed as \citep[e.g.][]{1979rpa..book.....R}
\begin{equation}
  \label{eq:bremssing}
  \frac{dW}{dE_\nu  dV dt}=\frac{16 \alpha_\textrm{f}r_\textrm{e}^2 c^2
  }{3u_\textrm{rel}}n_\textrm{e} n_\textrm{p}\log \left(\frac{b_\textrm{max}}{b_\textrm{min}}\right)\;,
\end{equation}
where $\alpha_\textrm{f}$ is the fine-structure constant,
  $r_\textrm{e}$ is the classical electron radius, $c$ the speed of light, 
$u_\textrm{rel}$ the relative
velocity of the ion with respect to the electron, and $n_\textrm{e}$ and $n_\textrm{p}$ are
the electron and ion number densities, respectively.

$b_\textrm{max}$ and $b_\textrm{min}$ are the maximum and
minimum value of the impact parameter. Through the definition of the Gaunt factor
\begin{equation}
  \label{eq:gaunt}
  g_\textrm{ff}(u_\textrm{rel},\omega)=\sqrt{3}/\pi\log (b_\textrm{max}/b_\textrm{min})\;,
\end{equation} 
where $\omega=2\pi \nu$ and $\nu$ is the frequency of the
bremsstrahlung photon, and subsequent use of quantum mechanical calculations of
$g_\textrm{ff}(u_\textrm{rel},\omega)$ we do not need to specify the
maximum and minimum impact parameters here and instead can use Gaunt
factors from the literature which already account for quantum mechanical
effects.

Substituting the Gaunt factor into (\ref{eq:bremssing}), we get
\begin{equation}
  \label{eq:bremssingrev}
  \frac{dW}{dE_\nu dV dt}=\frac{16\pi\alpha_\textrm{f}r_\textrm{e}^2 c^2
    }{3\sqrt{3}u_\textrm{rel}}n_\textrm{e} n_\textrm{p} g_\textrm{ff}(u_\textrm{rel},\omega)\;.
\end{equation}

For a one temperature plasma, the electrons are much faster (a
factor of $(m_\textrm{p}/m_\textrm{e})^{1/2}\approx 43$). Thus the protons can be
considered to be at rest and the relative velocity is dominated by the
speed of the electron. To get the electron-proton bremsstrahlung rate for this type of
plasma, it is sufficient to average the single proton emission rate over
the (assumed Maxwellian) velocity distribution of the electrons. Then
the energy of the photon is extracted from the kinetic energy of the
electron which slows, i.e. cools down. 

For a two-temperature plasma, however, the proton speed can become
comparable or even exceed the electron speed. Hence we need to
consider that and write the relative speed as 
\begin{equation}
  \label{eq:relative}
  u_\textrm{rel}=\sqrt{u_\textrm{e}^2+u_\textrm{p}^2-2u_\textrm{e}u_\textrm{p}\cos\theta}\;,
\end{equation}
where $u_\textrm{e}$ and $u_\textrm{p}$ are the electron and proton speed, respectively,
and $\theta$ the angle between the two speed vectors. 

For an isotropic Maxwellian distribution for the electrons and protons we
have the probability $dP_\textrm{e}$ for an electron to have a speed
between $u_\textrm{e}$ and $u_\textrm{e}+du_\textrm{e}$ 
\begin{equation}
    dP_{e}=\sqrt{\frac{2}{\pi}} \left(\frac{m_{e}}{kT_{e}}\right)^{3/2} u_\textrm{e}^2
  \exp\left(-\frac{m_{e}u_\textrm{e}^2}{2k T_{e}}\right)
  du_\textrm{e}\;,\label{eq:maxwelliane}
\end{equation}
and the corresponding probability
$dP_\textrm{p}$ for a proton to have a speed between $u_\textrm{p}$ and $u_\textrm{p}+du_\textrm{p}$
\begin{equation}
  dP_{p}=\sqrt{\frac{2}{\pi}} \left(\frac{m_{p}}{kT_{p}}\right)^{3/2} u_\textrm{p}^2
  \exp\left(-\frac{m_{p}u_\textrm{p}^2}{2k T_{p}}\right) du_\textrm{p}\label{eq:maxwellianp}\;,\\
\end{equation}
where $k$ is the Boltzmann constant.
Note that $\int dP_{e}$ and $\int dP_{p}$ are normalised to unity.

For the total emissivity of the electron-proton bremsstrahlung we have to
average over $dP_\textrm{e}$ and $dP_\textrm{p}$, respectively. We have to calculate,
using eqns.~(\ref{eq:bremssingrev}), (\ref{eq:relative}),
(\ref{eq:maxwelliane}) and (\ref{eq:maxwellianp})
\begin{equation}
  \label{eq:emissivity}
  \epsilon_\nu=\int_{u_\textrm{min,p}}^\infty \int_{u_\textrm{min,e}}^\infty
  \frac{dW}{dE_\nu dV dt} dP_\textrm{e} dP_\textrm{p}\;,
\end{equation}
where $u_\textrm{min,e}=(2 E_\nu/m_\textrm{e})^{1/2}$ and $u_\textrm{min,p}=(2 E_\nu/m_\textrm{p})^{1/2}$ accounts for the photon
discreteness effect. 
The result then still depends on the angle $\theta$ to lie between the velocities
$u_\textrm{e}$ and $u_\textrm{p}$. We then need to integrate over the probability for an
angle $\theta$ to occur. For an assumed isotropic distribution of angles, the probability for an angle between $\theta$ and
$\theta+d\theta$ is
\begin{equation}
  \label{eq:probtheta}
  dP_\theta=\frac{1}{2}\sin \theta d\theta\;.
\end{equation}
The angle averaged integral can be written in the form 
\begin{equation}
\begin{split}
\epsilon_\nu&=\frac{16 \alpha_\textrm{f}r_\textrm{e}^2 c^2}{3\sqrt{3} }n_\textrm{e}n_\textrm{p}\bar g_\textrm{ff}
\left(\frac{m_\textrm{e}m_\textrm{p}}{k^2T_\textrm{e}T_\textrm{p}}\right)^{3/2}\\
%&\int_0^\pi \int_{u_\textrm{p,min}}^\infty
%\int_{u_\textrm{e,min}}^\infty
%\frac{u_\textrm{e}^2u_\textrm{p}^2\exp\left(-\frac{m_\textrm{e}u_\textrm{e}^2}{2kT_\textrm{e}}-\frac{m_\textrm{p}u_\textrm{p}^2}{2kT_\textrm{p}}\right)\sin
%  \theta du_\textrm{e}du_\textrm{p}d\theta}{\sqrt{u_\textrm{e}^2+u_\textrm{p}^2-2u_\textrm{e}u_\textrm{p}\cos\theta}}\;.
&\cdot \int_0^\pi \int_{u_\textrm{min,p}}^\infty
\int_{u_\textrm{min,e}}^\infty
\frac{u_\textrm{e}^2u_\textrm{p}^2}{u_\textrm{rel}} \exp\left(-\frac{m_\textrm{e}u_\textrm{e}^2}{2kT_\textrm{e}}-\frac{m_\textrm{p}u_\textrm{p}^2}{2kT_\textrm{p}}\right)\sin
  \theta du_\textrm{e}du_\textrm{p}d\theta\;.
\end{split}
\end{equation}

This finally has to be integrated over the photon energy to get the
bremsstrahlung rate. For the energy and temperature dependent Gaunt
factor, $\bar g_\textrm{ff}$, we use  
\citet{1998MNRAS.300..321S}. Thus we do not longer need to justify our
choice for the maximum and minimum impact parameter. Higher precision
values are now hidden in the tabulated values of $\bar g_\textrm{ff}$,
taken from \citet{1998MNRAS.300..321S}. 

The integral over the electron and proton velocities has to be taken
carefully. As long as $u_\textrm{e}>u_\textrm{p}$, the electron-proton bremsstrahlung is
mainly caused by the kinetic energy change of the
electron. For $u_\textrm{e}<u_\textrm{p}$, the electron-proton bremsstrahlung is caused
by the kinetic energy change of the proton,
as then the electrons do not have much energy to put in the bremsstrahlung. 
The electron-proton bremsstrahlung rate, however, is then reduced by a factor
$m_\textrm{e}/m_\textrm{p}$, correcting for centre of mass effects.

In our calculations, we divided the velocity integrals accordingly and
get the contributions of protons and electrons to the electron-proton bremsstrahlung rate.
 
We give the corresponding analytic results for two limiting cases,
\[
  \epsilon_{\nu}\left(T_\textrm{e}\gg  T_\textrm{p}\frac{m_\textrm{e}}{m_\textrm{p}}\right)=\frac{16}{3}
    \alpha_\textrm{f}r_\textrm{e}^2 c^2n_\textrm{e}n_\textrm{p}\bar g_\textrm{ff}
\sqrt{\frac{2\pi m_\textrm{e}}{3 kT_\textrm{e}}}\exp\left(-\frac{E_\nu}{kT_\textrm{e}}\right)
\]
and 
\[
  \epsilon_{\nu}\left(T_\textrm{e}\ll  T_\textrm{p}\frac{m_\textrm{e}}{m_\textrm{p}}\right)=\frac{16}{3}\frac{m_\textrm{e}}{m_\textrm{p}}
    \alpha_\textrm{f}r_\textrm{e}^2 c^2 n_\textrm{e}n_\textrm{p}\bar g_\textrm{ff}
\sqrt{\frac{2\pi m_\textrm{p}}{3 kT_\textrm{p}}}\exp\left(-\frac{E_\nu}{kT_\textrm{p}}\right)\;,
\]
where the first result includes a one-temperature plasma, where
the protons are virtually at rest and the second corresponds to a two-temperature
plasma, where only the protons contribute to the electron-proton
bremsstrahlung and the electrons are virtually at rest.

Energy integrated this leads to 
\begin{equation}
  \epsilon\left(T_\textrm{e}\gg
  T_\textrm{p}\frac{m_\textrm{e}}{m_\textrm{p}}\right)=\frac{16}{3} \alpha_\textrm{f}
    r_\textrm{e}^2 c^2n_\textrm{e}n_\textrm{p}\tilde g_\textrm{ff}
\sqrt{\frac{2\pi }{3 } m_\textrm{e} k T_\textrm{e}}\label{eq:qbrp2}
\end{equation}
and
\begin{equation}
  \epsilon\left(T_\textrm{e}\ll
  T_\textrm{p}\frac{m_\textrm{e}}{m_\textrm{p}}\right)=\frac{16}{3}\frac{m_\textrm{e}}{m_\textrm{p}} \alpha_\textrm{f}
    r_\textrm{e}^2c^2 n_\textrm{e}n_\textrm{p}\tilde g_\textrm{ff}
\sqrt{\frac{2\pi }{3}m_\textrm{p} k T_\textrm{p}}\;.\label{eq:qbrp}
\end{equation}
Here, $\tilde g_\textrm{ff}$ is the frequency, i.e. photon energy, averaged Gaunt factor
(approximately $1.2\dots 1.5$)

The results for the one-temperature plasma reproduces the well known
result from the literature \citep[e.g.][]{1979rpa..book.....R}. For
the two-temperature plasma the electron-proton bremsstrahlung rate
depends on the temperature of the protons, accordingly. In either case the formulae only
depend on the temperature of the species which mainly cause
bremsstrahlung. The general dependence on the temperature is the same
in both limiting cases (proportional to $T^{-1/2}\exp\left(-E_\nu/(kT)\right)$ 
in the energy-dependent and $T^{1/2}$ in the
energy integrated case).

\begin{figure}
  \centering
  \includegraphics[width=0.34\textwidth,angle=-90]{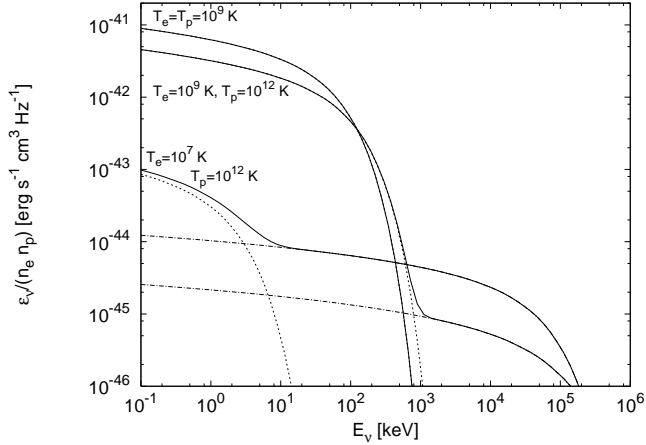}
  \caption{Electron-proton Bremsstrahlung emissivity $\epsilon_\nu$ of a optically
    thin gas for different combinations of electron and proton
    temperatures. The total emissivity is plotted as a solid line, the
    contribution of electrons as a dotted line and the proton
    contribution as a dash-dotted line. Note the different relative contributions of proton
    and electron bremsstrahlung for the same proton temperature, but
    different electron temperature. }
  \label{fig:spec}
\end{figure}

\begin{figure}
  \centering
  \includegraphics[width=0.34\textwidth,angle=-90]{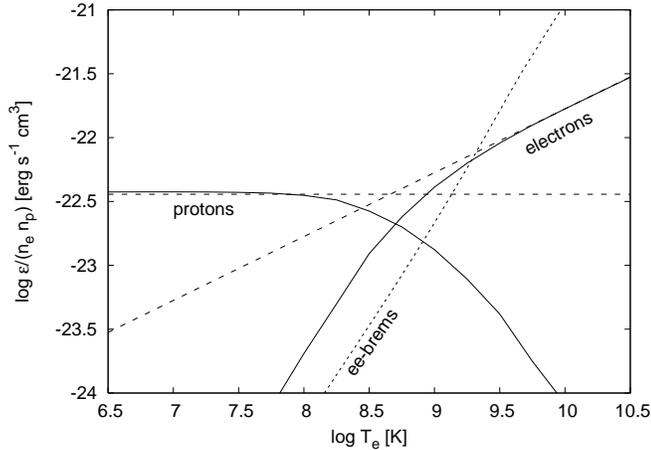}
  \caption{Electron-proton Bremsstrahlung emissivity of a optically
    thin gas for different electron temperatures at a proton
    temperature of $T_\textrm{p}=10^{12}$~K. For electron temperatures lower than
    $m_\textrm{e}/m_\textrm{p}T_\textrm{p}\approx 5.4\cdot 10^8$~K the contributions of the
    protons dominate (Then the electron-proton bremsstrahlung rate is
    essentially independent of the electron temperature),
    while for larger electron temperatures the
    classical result is reproduced. The long-dashed lines indicate the
    limiting cases calculated in eq.~(\ref{eq:qbrp2}) and
    (\ref{eq:qbrp}), respectively, while the short-dashed line gives
    the electron-electron Bremsstrahlung rate of
    \citet{1982ApJ...258..335S}. Note that for electron temperatures
    in excess of approx. $3\cdot 10^9$ K the electrons are
    relativistic and the plotted bremsstrahlung rate may be a very
    crude approximation in this regime.}
  \label{fig:contrib}
\end{figure}

We show sample electron-proton bremsstrahlung spectra for a gas at different electron
and proton temperatures in Fig.~\ref{fig:spec}. Note that the electron
component of the electron-proton bremsstrahlung shows a high-energy cut-off at lower energies
than the proton component. Although the magnitude of electron-proton
bremsstrahlung caused by the kinetic energy of protons is always lower
than that of bremsstrahlung created by the kinetic energy of electrons
for the cases considered here, the total, energy integrated
bremsstrahlung contribution of protons can be dominant due to the higher high
energy cut-off. This effect is shown in Fig.~\ref{fig:contrib} where
we plot the energy integrated electron-proton
bremsstrahlung rate for protons and electron as a function of electron
temperature. We also plot the corresponding electron-electron
bremsstrahlung rate, taken from \citet{1982ApJ...258..335S}.

The results for the electron-proton bremsstrahlung are only valid for non-relativistic protons and
electrons. However a first-order estimate of the relativistic correction can
be made by multiplying the energy integrated bremsstrahlung rate
with $\theta_\textrm{p}^{1/2}$ and $\theta_\textrm{e}^{1/2}$ in the relativistic
case ($\theta_\textrm{e}=kT_\textrm{e}/(m_\textrm{e} c^2)$ and $\theta_\textrm{p}=kT_\textrm{p}/(m_\textrm{p} c^2)$). This leads to a steepening of
the temperature dependence from $T^{1/2}$ in the non-relativistic to
$T$ in the relativistic energy integrated
case {\cb as indicated by \citet{1982ApJ...258..335S} for $\theta_\textrm{e}\gg
  1$. \citet{1980ApJ...238.1026G,1981ApJ...243..677G} finds a
  first-order correction factor of $(1+11/24\theta_\textrm{e})$ for $\theta_\textrm{e}\ll 1$.} 

Fig.~\ref{fig:contrib} shows that for electron
temperatures in excess of $10^9$ K electron-electron bremsstrahlung is
the dominant emission process. Hence a relativistic treatment of
electron-proton bremsstrahlung may not be
necessary. \citet{1975ZNatA..30.1546H} however shows that although
electron-electron bremsstrahlung dominates the emission the electron-proton
bremsstrahlung still contributes half of the electron-electron
bremsstrahlung rate.

\section{Conclusions}\label{sect:conclusions}

We have recalculated electron-proton bremsstrahlung for a
two-temperature plasma. Owing to the increasing importance of the
proton speed relative to the electron speed for $T_\textrm{e}< T_\textrm{p}$, the protons
can contribute significantly to the electron-proton bremsstrahlung. They dominate
the electron-proton Bremsstrahlung losses for electron temperatures lower than
$m_\textrm{e}/m_\textrm{p}\approx 1/1836$ times the proton temperature. 

While our results are strictly valid only in the
non-relativistic regime, we give a crude extrapolation for the
relativistic case. The work presented here should only be considered
as exploratory work to outline the influence of a two-temperature
plasma on the electron-proton bremsstrahlung. If calculated for both the non-relativistic and
relativistic regime, one needs to account for the relativistic
kinematics and consider the \citet{betheheitler} cross section,
adjusted for the effect presented here. While it is certainly needed
and desirable, such a treatment is beyond the scope of this paper. The
apparent dominance of electron-electron bremsstrahlung for electron
temperatures in excess of $10^9$ K, however, does not make it that necessary.

The mechanism presented here, seen in the context of Coulomb
collisions between electrons and protons, is already well known in plasma physics. The NRL Plasma
Formulary \citep{NRL_FORMULARY_06}, a standard reference in plasma
physics for more than 25 years, gives
a formula for the Coulomb collision rate which depends only on the
electron temperature for $T_\textrm{e}>m_\textrm{e}/m_\textrm{p} T_\textrm{p}$. This is the classical
\citet{1962pfig.book.....S} result. For higher proton temperatures,
the electron temperature dependence weakens and the collision rate
only depends on the proton temperature owing to their larger speed
compared to the electrons. This effect is seen as well in the work by
\citet{1983MNRAS.202..467S} who derives the Coulomb collision rate in
a fully relativistic framework. Similar to our extrapolation
for the electron-proton bremsstrahlung rate,
they find a change in the exponent of the temperature dependence of
$+1/2$, i.e. from $T^{-3/2}$ to $T^{-1}$. 

The importance of the effect presented in this paper needs to be
examined by comparing the characteristic timescales for Coulomb
collisions to equilibrate electron and proton temperature and the
Bremsstrahlung timescale. While it may not be of that strong influence
for the energetics of the two-temperature plasma, it could have an
observable signature due to the higher high-energy cut-off for the
Bremsstrahlung created by the kinetic energy of the protons, if it is
not hidden behind some more important emission mechanisms at these
energies.

\begin{acknowledgements}
  MM thanks J.E. Pringle for encouraging this work and helpful
  discussions. MM acknowledges
  support from PPARC and useful suggestions by the anonymous referee.
\end{acknowledgements}

\end{document}